# Amplitude modulation of wind turbine noise

Rufin Makarewicz, Roman Gołębiewski

Institute of Acoustics
A.Mickiewicz University
61-614 Poznan, Umultowska 85,
Poland
e-mail: roman_g@amu.edu.pl

**Abstract**

Due to swish and thump amplitude modulation, the noise of wind turbines cause more annoyance than other environmental noise of the same average level. The wind shear accounts for the thump modulation (van den Berg effect). Making use of the wind speed measurements at the hub height, as well as at the top and the bottom of the rotor disc (Fig.1), the non-standard wind profile is applied. It causes variations in the A-weighted sound pressure level, $L_{pA}$. The difference between the maximum and minimum of $L_{pA}$ characterizes thump modulation (Fig.2).

# 1. INTRODUCTION

Due to amplitude modulation, the wind turbine causes more annoyance than other environmental noise of the same average level [1-3]. Close and far away from the turbine, this modulation is referred as a "swish" and "thump", respectively [4,5]. The latter is evident at times of high wind speed shear (Sect.2). Thumping could be accompanied by swishing but swishing is not usually accompanied by thumping. In Ref.[6] measurements of amplitude modulation are discussed.

The swish is explained, among other things, in Ref.[7]: due to the strict directivity of sound generated by trailing edge and convective amplification, the noise emitted to the ground is produced mainly during the downward movement of the blades. Ref.[8] explains thumping by the same mechanisms. The alternative explanation of thump is based on the van den Berg effect [9-11]: modulation comes from both the vertical wind speed shear and horizontal wind direction shear (twists).

There are point source models of wind turbine noise which predict the A-weighted sound pressure level, $L_{pA}$ (e.g. see Refs.[12-18]). If $r$ is the distance between the turbine hub and the receiver, then

$$L_{pA} \approx L_{WA} - 10\log\left(\frac{2\pi r^2}{l_o^2}\right) - \Delta L(r), \; l_o = 1m. \tag{1}$$

Here $\Delta L(r)$ describes air attenuation, ground effect, and other propagation phenomena. In Sect 3 it is shown that the speed wind shear accounts for the modulation of the A-weighted sound power level,

$$L_{WA}(t) = 10\log\left\{\frac{W_A^{(o)} \cdot F(t)}{W_o}\right\}, \; W_o = 10^{-12}[W], \tag{2}$$

and consequently, accounts for the $L_{pA}$ time variations (Fig.2). The time function $F(t)$ varies with the frequency of the three blades rotation. The tip blade at the top and at the bottom of the rotor disc correspond to two extrema of the A-weighted sound pressure level, $L_{pA}^{(1)}$ and $L_{pA}^{(2)}$. Far away from a turbine the $L_{pA}$ time variations (Fig.2) are heard as a thump, which is characterized by the difference (Eqs.1,2),

$$\Delta L_{pA} = L_{pA}^{(1)} - L_{pA}^{(2)} = 10 \log \frac{F_1}{F_2}. \tag{3}$$

In order to find $\Delta L_{pA}$ the explicit form of $F(t)$ is needed (Sect.3).

## 2. WIND SPEED PROFILE

To determine the sound power emitted by the rotor disc (Fig.1), the wind profile $V(z)$ between its bottom, $z = h - l$, and top, $z = h + l$, has to be known. The standard wind profile power law can be written as follows (Fig.3):

$$V(z) = V(h) \cdot \left(\frac{z}{h}\right)^{\beta}, \quad 0 < z < \infty, \tag{4}$$

where $V(h)$ denotes the wind speed at the hub height, $h$. The measurements of the wind speed at two heights, e.g., $V(h+l)$ and $V(h)$, yield the standard wind shear exponent,

$$\beta = \frac{\log[V(h+l)/V(h)]}{\log[1+h/l]}. \tag{5}$$

The surface roughness and air stability influence the value of $\beta$. In urban and rural areas, $0.15 < \beta < 0.3$ and $0.07 < \beta < 0.55$, respectively (Fig.3.a) [19]. However Bowdler [20] has found that at the low wind speed of $V \approx 1 m/s$ (measured at the height $z = 10m$), the standard wind shear exponent could be quite large: $\beta \approx 4$. Note that inequality $\beta > 1$ implies a relatively big difference between the wind speeds at the top and bottom of the rotor disc, $V(h+l) - V(h-l)$ (red line on Fig.3b).

If the third measurement of the wind speed, $V(z_1)$, is performed at the height $z_1$, then the generalization of Eq.(4) becomes,

$$V(z) = [V(h) - V(z_1)] \cdot \left(\frac{z - z_1}{h - z_1}\right)^\gamma + V(z_1), \qquad (6)$$

where $\gamma > 0$ denotes the non-standard wind shear exponent. The rotor disc radiates noise (Fig.1). Therefore the wind speed measurement $V(z_1)$ is recommended at the height of its bottom, $z_1 = h - l$. Accordingly, we write the non-standard wind profile as (Ref.10, Fig.4),

$$\hat{V}(z) = V(h) \cdot \left[a + (1-a)\left(1 + \frac{z-h}{l}\right)^\gamma\right], \quad h - l < z < h + l. \qquad (7)$$

The wind parameter $a$ and the non-standard wind shear exponent $\gamma$ can be calculated from,

$$a = \frac{V(h-l)}{V(h)}, \quad \gamma = \frac{10}{3}\log\frac{V(h+l) - V(h-l)}{V(h) - V(h-l)}, \qquad (8)$$

where $0 < a \leq 1$ and $0 \leq \gamma < \infty$. Function (7) holds true when the wind speed grows with height, $V(h-l) \leq V(h) \leq V(h+l)$. The values of $a$ and $\gamma$ are related to each other by real wind profiles, $V(z)$. For example, the uniform wind inside the rotor disc, $V(h-l) = V(h) = V(h+l)$, corresponds to $a = 1$ and $\gamma = 0$.

## 3. NOISE GENERATION

To describe the process of noise generation, we begin from the empirical relationship between the A-weighted sound power level,

$$L_{WA} = 10\log\frac{\langle W_A \rangle}{W_o}, \qquad (9)$$

and the wind speed at the hub height, $V(h)$. Here $\langle W_A \rangle$ expresses the time average A-weighted sound power. The measurement data from Ref.[24-26] can be described by,

$$L_{WA} = 10n \cdot \log\frac{V(h)}{V_o} + B, \quad V_O = 1 m/s, \qquad (10)$$

where the sound power parameter, $1 < n < 4$, is a function of the turbine type and the power mode. To explain the above dependence, we assume that the bulk of the blade noise is produced by its tip at the distance $l$ from the hub (Fig.5, [26]).

If $N$ [rps] denotes the rotational speed, then the momentary height of a tip is a function of time $t$,

$$z(t) = h + l\cos\Phi \text{ where } \Phi = 2\pi N l \cdot t. \qquad (11)$$

When the blade tip moves periodically between $z = h + l$ (rotor disc top) and $z = h - l$ (rotor disc bottom), it encounters the wind speed given either by (Eqs. 4,11),

$$V[z(t)] = V(h) \cdot \left\{1 + \frac{l}{h}\cos\Phi\right\}^{\beta}, \qquad (12)$$

or by (Eqs.7,11),

$$\hat{V}[z(t)] = V(h) \cdot \left[a + (1-a)(1+\cos\Phi)^{\gamma}\right], \qquad (13)$$

where the angle $\Phi = 2\pi N l \cdot t$ (Fig.5). The wind interaction with the moving blade brings about the inflow-turbulence noise [7,8], which grows with the wind speed. So the time varying A-weighted power of sound from a single blade can be written as,

$$W_A = q \cdot \left[\frac{V(z)}{V_o}\right]^n, \quad V_o = 1[m/s]. \tag{14}$$

Here $q$ represents the unknown coefficient (see below) and $n$ is the known empirical sound power parameter (Eq.10). Due to time variations of the tip height $z$ (Eq.11), the instantaneous A-weighted sound power from three blades becomes a product,

$$W_A(t) = W_A^{(o)} F(t), \tag{15}$$

where

$$W_A^{(o)} = q \cdot \left[\frac{V(h)}{V_o}\right]^n, \tag{16}$$

and the explicit form of $F(t)$ is discussed below for standard and non-standard wind profiles.

Standard wind profile

Fig.3 shows the standard wind profile $V(z)$. Due to three being three blades (Fig.1) the modulation function $F(t)$ (Eqs.12,14,15) is a sum of three terms,

$$F(t) = g(\Phi) + g(\Phi - 2\pi/3) + g(\Phi + 2\pi/3), \tag{17}$$

where

$$g(\Phi) = \left[1 + \frac{l}{h}\cos\Phi\right]^{\beta n}, \quad \Phi = 2\pi N l \cdot t. \tag{18}$$

Mindful the time period, $T = 1/N$ (Eq.11), the integrals,

$$\langle W_A \rangle = \frac{1}{T}\int_0^T W_A dt = \frac{1}{2\pi}\int_0^{2\pi} W_A(\Phi) d\Phi, \tag{19}$$

yield the time average value of $W_A$ (Eq.15). Then the combination of Eqs.(15-19) results in,

$$\langle W_A \rangle = 3q \langle g \rangle \cdot \left[ \frac{V(h)}{V_o} \right]^n . \tag{20}$$

The average value of $\langle g \rangle$ (Eqs.18,19),

$$\langle g \rangle = \frac{1}{2\pi} \int_0^{2\pi} \left[ 1 + \frac{l}{h} \cos \Phi \right]^{\beta n} d\Phi , \tag{21}$$

increases with the ratio of the blade length over the hub height, $l/h$ (Fig.1). Finally, the time average A-weighted sound power $\langle W_A \rangle$ (Eq.20) and the A-weighted power level (Eq.9) combine into a relationship (10) with the constant, $B = 10 \log [3q \langle g \rangle]$. Then using the measured value of $B$ (Eq.10, Refs.[24-26]), one finds the unknown coefficient $q$ (Eq.14).

While the blade tip rotates (Eq.11), the periodical changes of its height $z(t)$ yields the periodic variations in the encountered wind speed $V[z(t)]$ (Eq.12). By using Eqs.(17) and (18) one arrives at the modulation function $F(\Phi)$, which stems from the standard wind profile power law, $V(z)$ (Eq.4),

$$F(\Phi) = \left[ 1 + \frac{l}{h} \cos \Phi \right]^{\beta n} + \left[ 1 + \frac{l}{h} \cos \left( \Phi - \frac{2\pi}{3} \right) \right]^{\beta n} + \left[ 1 + \frac{l}{h} \cos \left( \Phi + \frac{2\pi}{3} \right) \right]^{\beta n} , \tag{22}$$

where $\Phi = 2\pi Nl \cdot t$.

As an example, Fig.6 shows the plots of $g(\Phi)$, $g(\Phi - 2\pi/3)$, and $g(\Phi + 2\pi/3)$ (Eq.18) for the ratio $l/h = 0.5$ (Fig.1), a relatively large wind shear exponent, $\beta = 1.25$ (Fig.3.b), and the sound power parameter, $n = 4$ (Eq.10). For any $l/h$, $\beta$, and $n$, two extrema of $F(\Phi)$ occur at $\Phi = 0$ and $\Phi = \pi/3$,

$$F_1 = g(0) + g(-2\pi/3) + g(+2\pi/3), \quad F_2 = g(\pi/3) + g(-\pi/3) + g(\pi). \tag{23}$$

All in all, for the standard wind profile power law (Fig.3, Eqs.4,5), the thump characteristic can be calculated from (Eqs.3,22,23),

$$\Delta L_{pA} = 10 \log \frac{\left[1 + \frac{l}{h}\right]^{\beta n} + 2 \cdot \left[1 - \frac{l}{2h}\right]^{\beta n}}{\left[1 - \frac{l}{h}\right]^{\beta n} + 2 \cdot \left[1 + \frac{l}{2h}\right]^{\beta n}}, \quad \beta n > 2. \tag{24}$$

The dependence of $\Delta L_{pA}$ on the product $\beta n$, for ratios $l/h = 0.3$, 0.5, and 0.7, is plotted in Fig.7. Note that the thump modulation begins with $\beta n = 2$ and increases with $l/h$.

Non-standard wind profile

For a non-standard wind profile power law (Fig.4, Eqs.6,7), the modulation function takes the form (Eqs. 13,14,17),

$$\hat{F}(t) = \hat{g}(\Phi) + \hat{g}(\Phi - 2\pi/3) + \hat{g}(\Phi + 2\pi/3), \tag{25}$$

where

$$\hat{g}(\Phi) = \left[a + (1-a)(1+\cos\Phi)^{\gamma}\right]^n, \quad \Phi = 2\pi Nt. \tag{26}$$

The measured wind speeds, $V(h-l) < V(h) < V(h+l)$, give both the wind parameter $a$ and the non-standard wind shear exponent, $\gamma$ (Eq.8). Similarly to the case of the standard wind shear, two extrema of the modulation function $\hat{F}(t)$ are determined by (Eq.23),

$$\hat{F}_1 = \hat{g}(0) + \hat{g}(-2\pi/3) + \hat{g}(+2\pi/3), \quad \hat{F}_2 = \hat{g}(\pi/3) + \hat{g}(-\pi/3) + \hat{g}(+\pi). \tag{27}$$

Ultimately, the thump characteristic becomes (Eqs.3,26,27),

$$\boxed{\Delta L_{pA} = 10\log \frac{[a+(1-a)\cdot 2^{\gamma}]^n + 2\cdot[a+(1-a)\cdot 2^{-\gamma}]^n}{a^n + 2\cdot\left[a+(1-a)\left(\frac{3}{2}\right)^{\gamma}\right]^n}}, \tag{28}$$

where $\Delta L_{pA} > 1$ dB. The concurrent measurements of the A-weighted sound power, $L_{WA}$, and the wind speed at the hub height, $V(h)$, yield the sound power parameter $n$ (Eq.10). To compute the values of *a* and $\gamma$ (Eq.8), three wind speed measurements $V(h-l) \leq V(h) \leq V(h+l)$ (Fig.4) are necessary. In Fig.8 one can find examples of the function $\Delta L(a,\gamma,n)$ (Eq.28).

## 4. CONCLUSIONS

This study concentrates on noise generation by the rotor disc of wind turbine, $h-l<z<h+l$ (Fig.1). The non-standard wind profile for the rotor disc, $\hat{V}(z)$ (Fig.4, Eqs.6,7), with coefficients $a$ and $\gamma$ (Eq.8) is derived from three wind speed measurements, $V(h-l) \leq V(h) \leq V(h+l)$. The thump characteristic is defined as the difference of the A-weighted sound pressure level, $\Delta L_{pA}$ (Fig.2, Eq.3). The model of $\Delta L_{pA}$ (Eq.28) estimation is based on the fact that the bulk of the sound energy is emitted by the blade tip [26]. As expected, the value of $\Delta L_{pA}$ increases with the sound power parameter, $n$ (Eq.10), and the non-standard wind shear exponent, $\gamma$ (Fig.8).

**Figure captions**

Fig.1. The wind turbine with blades of length $l$ and the hub at the height $h$.

Fig.2. Thumping comes from variations in the A-weighted sound pressure level between two extrema, $L_{pA}^{(1)}$ and $L_{pA}^{(2)}$.

Fig.3. Standard wind profile $V(z)$ (Eq.4) with wind shear exponents $\beta < 1$ (a) and $\beta > 1$ (b).

Fig.4. Non-standard wind profile $\hat{V}(z)$ (Eq.7) for the rotor disc, $h - l < z < h + l$, with the wind parameter $a$ and the wind shear exponent $\gamma$ calculated from Eq.(8).

Fig.5. A blade tip at the distance $l$ from the hub with the instantaneous angle, $\Phi = 2\pi N t$.

Fig.6. Three elements of the modulation function $F(\Phi)$ with $\Phi = 2\pi Nl \cdot t$ (Eq.22), for the ratio $l/h = 0.5$ (Fig.1), standard wind shear exponent, $\beta = 1.25$ (Fig.3.b), and the sound power parameter, $n = 4$ (Eq.10).

Fig.7. The thump characteristic $\Delta L_{pA}$ (Eqs.1,24) for the standard wind profile (Fig.3, Eq.4), with the ratios, $l/h = 0.3, 0.5,$ and $0.7$ (Fig.1).

Fig.8. The thump characteristic $\Delta L_{pA}$ (Eqs.1,28) for the non-standard wind profile (Fig.4, Eqs. 6,7) with ($a = 0.1$; $\gamma = 2.0$), ($a = 0.9$; $\gamma = 2.5$), ($a = 0.5$; $\gamma = 3.0$).

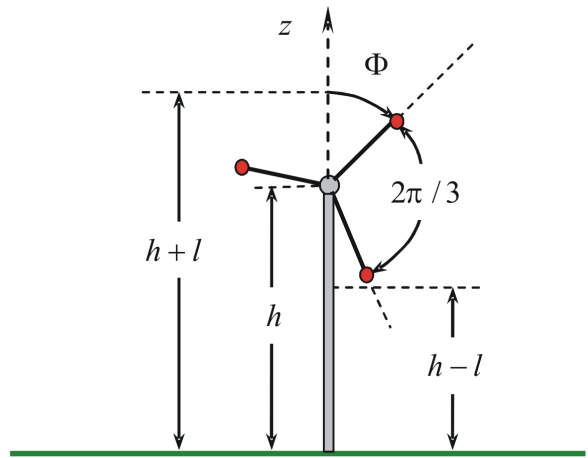

Fig.1

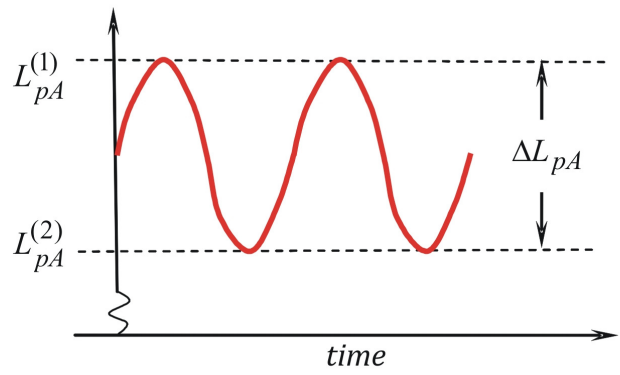

Fig.2

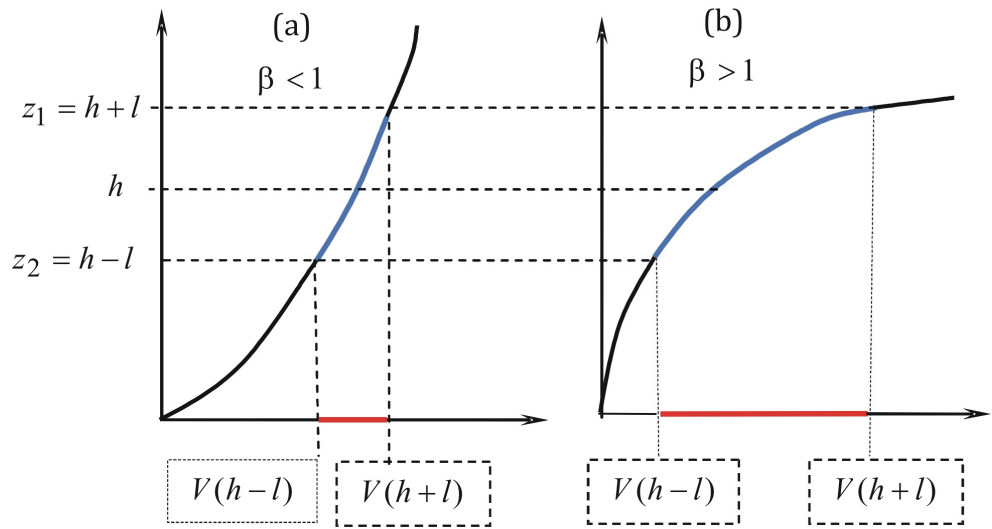

Fig.3

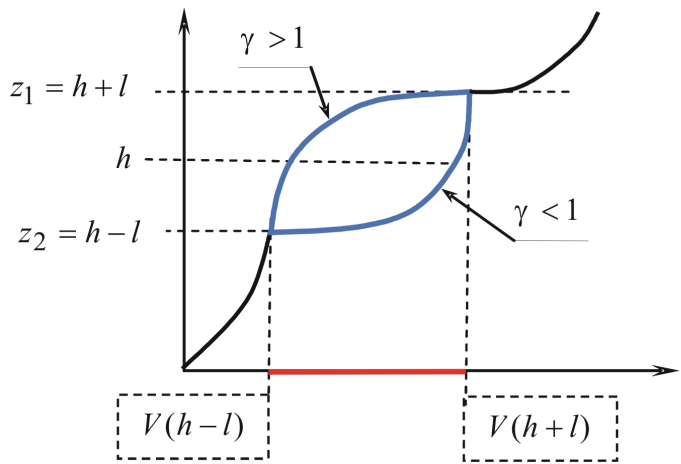

Fig.5

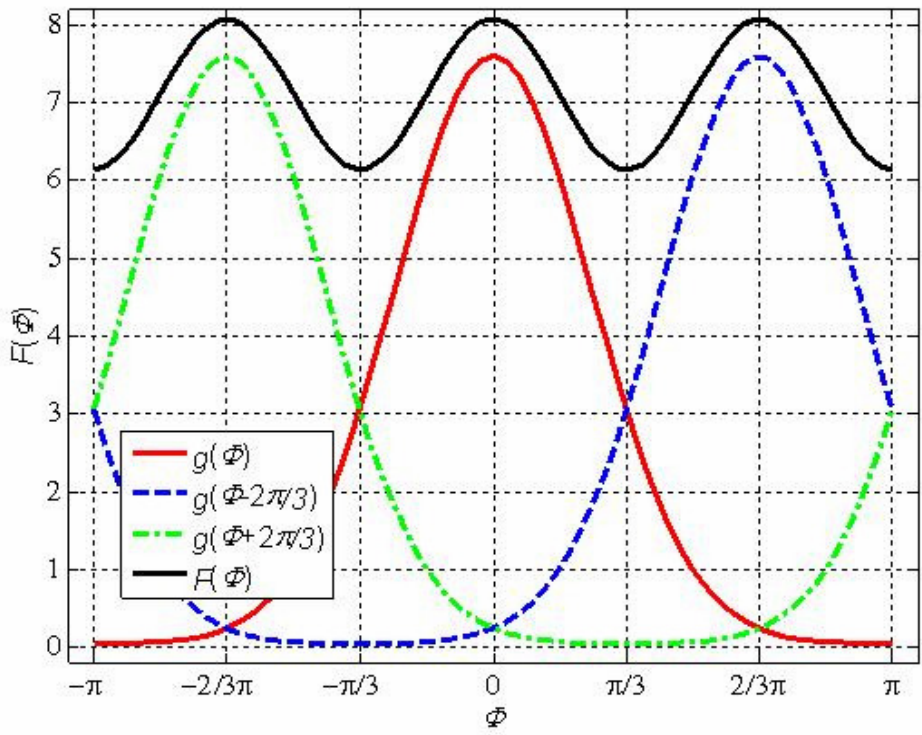

Fig.6

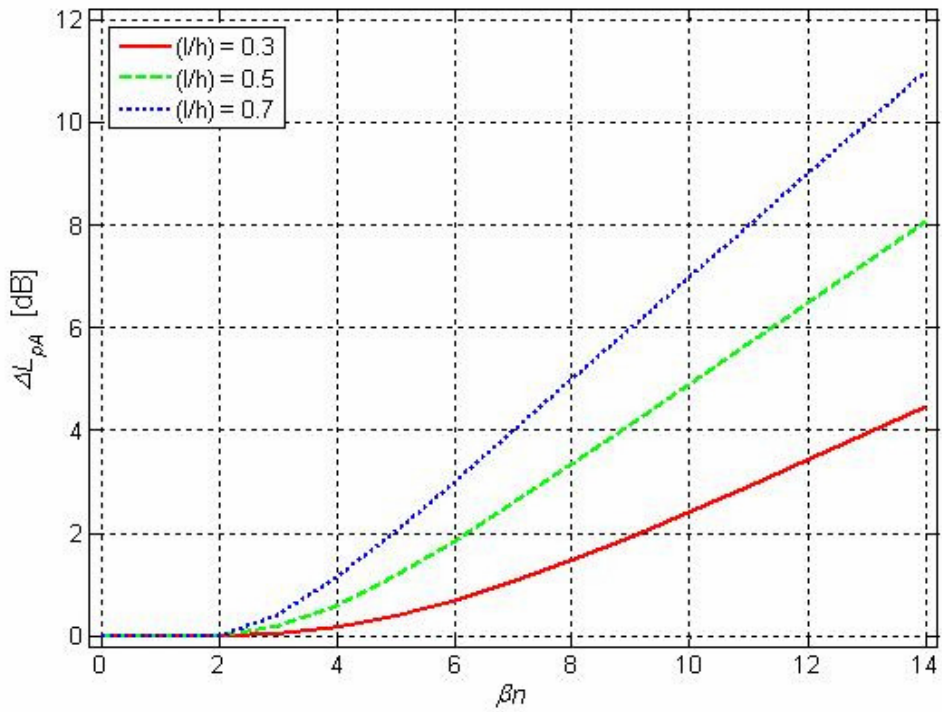

Fig.7.

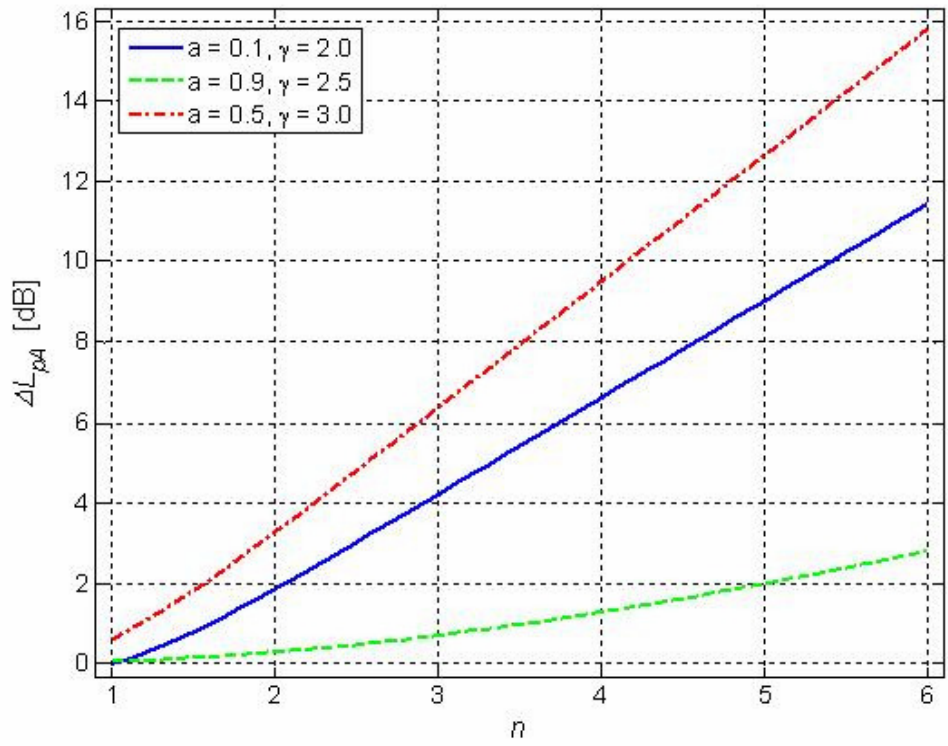

Fig.8